# Sustainable blockchain-enabled services: Smart contracts

Craig Wright
nChain , London , UK
craig@ncrypt.com

Antoaneta Serguieva
nChain , London , UK
antoaneta@ncrypt.com

*Abstract*—This chapter contributes to evolving the versatility and complexity of blockchain-enabled services through extending the functionality of blockchain-enforced smart contracts. The contributions include: (i) a method for automated management of contracts with hierarchical conditionality structures through an hierarchy of intelligent agents and the use of hierarchical cryptographic key-pairs; (ii) a method for efficient and secure matching and transfer of smart-contract underlyings (entities) among disparate smart contracts/subcontracts; (iii) a method for producing an hierarchy of common secrets to facilitate hierarchical communication channels of increased security in the context of smart contracts/subcontracts/underlyings; and (iv) a method for building secure and optimized repositories through distributed hash tables in the context of contracts/subcontracts/underlyings. These methods help providing services that allow both narrower and worldwide reach and distribution of resources. The longevity of the blockchain technology is achieved through continuous innovation. Blockchain-enabled services are potentially an efficient, secure, automated, and cost-effective alternative or complement to current service infrastructures in a range of domains (legal, medical, financial, government, IoT).

## I. INTRODUCTION

The acceleration of blockchain functionality aims at enabling complex services in a secure, efficient, and creatively automated manner. For a range of domains, blockchain-enabled services can provide viable alternatives or complements to existing service infrastructures, particularly to those currently underperforming or of unreliable security. The emerging research area of smart contracts plays a critical role in building the alternative and complementary infrastructures. The methods proposed in this paper are associated with smart contracts, and include innovative elements that have not been considered in the literature.

We can relate the proposed methods to existing open questions, as reviewed next:

▫ Challenges in validating and verifying smart contracts (SM) are recognized in [1], considering that SM may encode legal contracts written in natural language. The current paper addresses these challenges in Section II, and proposes a contract codification model along with a method for automated SM management, validation, and verification, enforced through the blockchain.

▫ The combination of the Internet-of-things (IoT) and blockchain is discussed in [2], and it is recognized that this combination can facilitate the sharing of services and resources through a marketplace of services among IoT devices. Therefore, corresponding solutions should be developed. The codification model introduced in the current paper translates a wide range of contracts. The concept of 'contract' is used in its broader meaning of structured control conditions. The method for automated management of such conditions, proposed in Section II next, is directly applicable and beneficial to emerging IoT services. The method proposed in Section III, for efficient and secure transfer of entities underlying smart contracts, also contributes to creating a marketplace of services between IoT devices. We support the view that the blockchain-IoT combination is powerful and can cause significant transformations across several industries.

▫ The potential of blockchain technology for innovating and transforming governmental processes is questioned in [3]. The conclusion there is that governmental processes would benefit most from the technology if blockchain applications are customized to fit process requirements. The secure repository of smart-contract templates, i.e. smart-process templates, which we propose in Section II here, contributes to implementing that conclusion. The repository mechanism involves each institution accessing the smart-templates repository to derive its institutional semi-templates and continuously amend them through reuse.

The remainder of the paper is organized as follows: Section II considers the automated management of blockchain-enforced smart contracts, Section III is focused on smart contracts' efficiency and security, and Section IV states the conclusions and indicates further research focus.

## II. AUTOMATED MANAGEMENT OF BLOCKCHAIN-ENFORCED SMART CONTRACTS

### A. Benefits

Smart contracts potentially extend the range of services facilitated through the blockchain technology, transform existing legal and financial infrastructures, and provide for emerging IoT services. Institutions, individuals, intelligent computing agents, or IoT devices play the role of counterparties in a smart contract. The method proposed here simplifies the management of such contracts, based on adapting, combining and implementing key components from [4][5][6][7]. The proposed innovation:

- allows contracts to be time-bound, condition-bound, open-ended, and rolling-over;
- introduces a security-enhanced control mechanism that permits or prohibits access to an off-chain contract repository in an intelligent manner;
- provides a formalism for translating any structured control conditions into a corresponding $contract\_model$ and its deterministic finite automation $DFA$;
- provides an intelligent-agent mechanism to follow and execute the embedded $DFA$ logic;
- allows for holding a secure public record of agents' code on the blockchain;
- introduces a mechanism for turning an unspent crypto-transaction $UTXO$ into an indication of the stage in executing the hierarchy of subcontracts (that allows control over different aspects of the overall contract to be partitioned) within a smart-contract structure;
- introduces a mechanism to hold a secure public record of current and past contracts on the blockchain, in a manner that allows automated determination of their validity and release of their details to authorized entities upon validation.

The initiation, stages of execution, and closure of a smart contract are recorded on the blockchain through the creation, broadcasting and recording of crypto-transactions. This allows verifying the current or past existence of a contract by looking up corresponding blockchain transactions. The stage of execution of an existing contract can also be verified, by looking up recorded transactions corresponding to the initiation or closure of its subcontracts. The logic of the structured control conditions from the $contract\_model$ is embedded within the locking/unlocking scripts of blockchain transactions, as well as within other transaction elements such as the nLockTime field. The contract's logic is enforced through individual actions and overall behavior of one or several intelligent-agent applications. For accountability and for reuse, parts of the codified overall behavior of an $agent$, or links to off-chain repositories where the code is stored, are also embedded within the locking/unlocking scripts of transactions recorded on the blockchain. The access to these repositories, as well as to repositories storing the contract documents and the $contract\_models$, is selective, partial, and secure. The access to the code or parts of it, and the access to the contract or its subcontracts, matches the requirements and character of the contract and the preferences of the multiple counterparties involved.

### B. Contract Model and Tokenisation

A repository of contracts can be implemented as a distributed hash table (DHT) [8] across storage resources within the network supporting a blockchain. A reference or link hash to a contract's entry in the repository is stored as metadata within a blockchain transaction, and serves as a DHT look-up key for referencing the contract from the blockchain. The use of a master encryption key and multiple sub-keys by each counterparty, as proposed in Section III.A next, allows for secure repository access of a counterparty to the contract or the subcontracts that this counterparty is authorized for. Auditing authorities are also provided with access corresponding to the scope of each audit. Consider the following example:

- A building company in England enters into a contract with multiple counterparties to deliver a new development. The contract has multiple subcontracts, and one of them addresses the issuance of a plans certificate, as required by the relevant regulation. One of the counterparties on this subcontract is the Building Control Department of a Local Authority. The Control Department has access to this and probably further subcontracts in the repository, but may not have access to subcontracts specifying remunerations for the pool of builders or other confidential details. Another subcontract of the building contract concerns the issuance of a final certificate, as required by regulation. A counterparty on this subcontract is an approved building inspector, and his secure access is defined by analogy with the former case. The building company and its auditors may have access to all of the subcontracts in the repository, and to all blockchain transactions enforcing the contract and its subcontracts. The auditing firm may not be a counterparty to any subcontract, and may access the repository after the completion of the contract. The auditors still are able to retrieve the relevant information and verify transactions throughout the past execution of the contract, and thus assess the performance of the building company.

The use of multiple encryption sub-keys also allows that trusted third parties may modify some of the conditionality and subcontracts of a stored contract. This translates into an amended behavior of the intelligent $agents$ enforcing the contract. The blockchain transactions, which the $agents$ create for the amended instantiation of the contract, include amended parameters in comparison with the transactions they created for a previous instantiation. For example, the renewal of a lease contract or the renewal of a rental contract may involve amended amounts and rates. Multiple encryption sub-keys (see Section III.A) further facilitate establishing a $common\_secret$ [5] for each pair of counterparties on each subcontract. A $common\_secret$ based encryption allows for a secure channel of communication between a pair of counterparties, when it is necessary to negotiate values of parameters related to a subcontract such as lease rates and rental amounts. Differing $common\_secrets$ between the same counterparties on different subcontracts provide for additional security.

Having considered, as prerequisite, the mechanism of a DHT repository for smart contracts, we now continue with the $contract\_model$. The list of its elements includes:
- a codification scheme that allows a complete description of any type of contract (structured control conditions), and is based on constructs such as XBRL, XML, JSON, etc.;
- a deterministic finite automaton $DFA$ 'translating' the contract logic and conditionality, where the $DFA$ can be fully defined within the codification scheme and consists of:
  - a set of parameters and indication where to source them;
  - a set of state definitions;
  - a set of transitions between the states, including the trigger for a transition and the rules followed during transition;

- rules definition table;
- definitions of the specific parameters for this instance of the contract;
- a 'compiler' converting the codification scheme into intelligent-agent code and crypto-transaction script.

The $DFA$ is the essential component of the $contract\_model$ and is implemented as an agent-based process. For complex contracts, the $DFA$ implementation involves a sequence of processes or parallel sequences of processes. Processes access off-chain resources, and/or monitor the values of off-chain and on-chain parameters, and/or create different blockchain transactions, under each conditionality step within the active contract and under different triggers and parameter values. Agent-based processes also send multisig transactions for signature by counterparties prior to broadcasting them, and/or communicate off-chain to inform counterparties or trusted third parties, and/or verify on-chain records related to the execution of past contracts. A $master\_agent$ can manage a hierarchy of $subordinate\_agents$ that carry out tasks defined in a smart contract. The $master\_agent$ controls, directs, monitors, and authorizes the activities of each $subordinate\_agent$, and also coordinates their activities. The $master\_agent$ and $subordinate\_agents$ communicate to execute the variety of tasks.

Having introduced the $contract\_model$, the first step in its implementation is to indicate the existence of a contract. The $master\_agent$ on this contract creates the first transaction $T$ associated with the contract, broadcasts it to the blockchain network, monitors when it is recorded on the blockchain and extracts its ID. Thus, the existence of the contract and the time when it became active are a permanent auditable record publicly available on the blockhcain, although the details of the contract may not be publicly available. The $master\_agent$ uses a pay-to-script-hash $P2SH$ address when creating transaction $T$. For such transaction to be spent, a recipient must provide a script matching the $P2SH$ script hash as well as data that makes the script evaluate to true. $P2SH$ is created using the contract metadata. After $T$, a number of further transactions follow that are associated with the contract and its subcontracts. They are created by the $master\_agent$ or by $subordinate\_agents$.

A range of these transactions involve tokenization. In the rest of this Section II.B, we introduce and extend tokenization mechanisms from [9][10]. Each $master\_agent$ or $subordinate\_agent$ has its own master key and sub-keys. The hierarchies of keys are used in combination with the tokenization mechanism and allow for contract structure of any complexity to be created and implemented, and for the related subcontracts and schedules to be confirmed, triggered, executed, and terminated. In this context, a token can non-exhaustively be used to represent and detail, in the form of a crypto-transaction, the transferable rights conferred by a specific contract or subcontract. The $token$ efficiently uses metadata comprising only three parameters:
- a number of units available overall, $total\_units$;
- a quantity of $transfer\_units$ to be transferred from a sender to at least one recipient;
- a $pegging\_rate$ for calculating a value for the $transfer\_units$ pegged to the crypto- currency.

Such $token$ can represent any type of transferable rights, and thus common algorithms are reused as parts of the codified behavior of different $agents$. The $token$ is either divisible or non-divisible, corresponding to the transfer of divisible or non-divisible rights. In the latter case, the value of the parameter $pegging\_rate$ is set to 0. For divisible rights, the tokenized value transferred in the transaction output is tied to the underlying crypto-currency amount via a non-zero $pegging\_rate$, and the transferred rights are specified in terms of a $pegging\_rate$. An example of divisible tokens are those transferring quantities of bearer shares, where a share is a percentage ownership (a pegging rate) of the company. An example of non-divisible tokens are those transferring bearer bonds, where a bond is redeemable for an exact amount of a fiat currency such as USD or GBP. If some smart contracts or their subcontracts involve issuing and selling bearer shares or bonds, then among the crypto-transactions being created during the implementation of the contracts are also transaction representing the transfer of tokenized quantities of shares or bonds.

Furthermore, the number of units available overall in the tokenized rights is either limited or unlimited. In the former case, the parameter $total\_units$ is fixed and always greater than 0. An example of limited units is the shared ownership of a race horse, such as $total\_units = 10$ and $pegging\_rate = 10\%$, or $total\_units = 25$ and $pegging\_rate = 4\%$. An example of unlimited overall units is the inventory of a product in a warehouse, as the inventory can be increased at any time and allow an increase in the tokenized amount of product units. Bearer shares are also an example of potentially unlimited units, due to the company being able to issue more shares. In some cases of unlimited units, the current total number of units does not matter for the transfer of ownership, and the value of parameter $total\_units$ in the token is set to 0. Such is the inventory example, where one unit is one instance of the transferable product – a T-shirt in the warehouse stock of T-shirts. In other cases, the current number of potentially unlimited total available units matters. As this number is variable, a $subordinate\_agent$ monitors it and identifies its correct value for each instantiation of such divisible token. In case of bearer shares, transferring tokenized ownership rights involves parameter values as follows:

$$total\_units = \text{current number of issued and non-redeemed shares}$$

$$pegging\_rate = \frac{1}{total\ units}\%$$

A $subordinate\_agent$ monitors the number of issued shares and the number of redeemed shares, and identifies the current value of $total\_units$.

A final point here is that the crypto-currency amount, which is attached to the output of a $token$ transaction, is arbitrary. Such transaction is only a facilitator of ownership transfer, and the true value of transferred rights is found through the metadata parameters. We introduce a meaningful use of a $token$'s output amount, under the proposed here specification of tokenization. In a divisible token, we link that amount to the $pegging\_rates$, when the token is split into several transaction outputs. Having considered the contract model and a relevant tokenization mechanism in this Section, the focus in the next Section II.C is on a contract's conditionality and subcontracts.

### C. Master Contracts' Conditionality and Subcontracts

A $master\_contract$ is interpreted here as remaining in effect, as long as there is a valid unspent transaction output $UTXO$ representing the existence of the contract. That unspent state is influenced and altered by the behavior of the $master\_agent$ and $subordinate\_agents$. Agents' behavior is controlled through conditions in the $master\_contract\ model$ that translate provisions and stipulations from the contract document. For example, a condition may involve that the contract expires when the values of some variables reach specified thresholds. Transactions associated with a contract are a permanent, unalterable public record of the contract's existence and current status. The termination of a contract is also recorded on the blockchain, as a spent output in a crypto-transaction. Anyone can use a software module to determine, from the blockchain, at what stage of its execution a contract is or whether it has been terminated.

In this context, a $subcontract$ is a contract that is directly related to an existing $master\_contract$, where the metadata in transactions associated with the $subcontract$ contain a hashed pointer or reference to the location of the $master\_contract$ within the DHT repository. The existence of a $subcontract$ is implemented, similarly to a $master\_contract$, as an $UTXO$ with a deterministic redeem script address. The $subcontract$ is interpreted as being completed when this $UTXO$ is spent. The steps used for creating the deterministic addresses in the $P2SH$ transactions associated with $subcontracts$, within a $master\_contract'$s conditionality structure, include the following:

- derive a new public sub-key using seed information;
- if an entry for a $subcontract$ does not exist in the repository of contracts, then create an entry so that:
  - the entry is a description of this $subcontract$ compliant with the codification scheme of the $contract\_model$ introduced in Section II.B;
  - this description includes a reference to the $master\_contract$ entry in the repository;
- once the $subcontract$ entry is created or if such entry already exists in the repository, then add the reference to this entry to the metadata of crypto-transactions associated with the $subcontract$;
  - these metadata may also include a reference to the $master\_contract$ entry;
- use the amended metadata to create $P2SH$ addresses.

A use-case for creating subcontracts is described in Table 1. The mechanism described at step *six* in Table 1 is also used to monitor further types of conditions within a given $master\_contract$. For example, if a contract is worth a Z amount of crypto-currency with $Z_1, \cdots, Z_k$ to be paid at checkpoints 1 through $k$, then this is implemented as a $master\_contract$ plus $subcontracts$. Each of the $subcontracts$ is marked as complete using the same or different signatories ($agents$, notaries, surveyors, brokers). Thus, a public record is maintained showing which of the conditions attached to the $master\_contract$ have been met and which are yet to be met. For $i \epsilon \{1, \cdots, k\}$, a $subordinate\_agent_i$ monitors the state of $subcontract_i$ and triggers payment $Z_i$, once the monitoring confirms that $subcontract_i$ is complete.

TABLE I. ISSUING A SUBCONTRACT BASED ON AN EXISTING CONTRACT

| Step | Details |
|---|---|
| one | The $master\_agent$ derives, using a seed value, a new public sub-key from its master public key used to create the $master\_contract$. The $master\_contract\_issuer$ derives, using the same seed, a new public sub-key from his master public key used for the $master\_contract$. The $master\_contract\_issuer$ can be an institution or an individual responsible off-chain for the $master\_contract$. The seed value is based on information about the $master\_contract$. Examples of appropriate seeds include:<br>-Transaction ID of the $UTXO$ published on the blockchain to indicate the existence of the $master\_contract$;<br>-Redeem script hash securing the $master\_contract$ and created by the $contract\_issuer$ or the $master\_agent$ in an $m$-of-$n$ multi-signature structure, where at least the public keys of the $contract\_issuer$ and the $master\_agent$ must be supplied to this script. Depending on the terms of the $master\_contract$, other signatures may also be required, including the signatures of a $subordinate\_agent$ and a $subcontract\_issuer$, where the $subcontract\_issuer$ has responsibilities for the $subcontract$ in the off-chain world. The number $n$ further includes the number of metadata blocks.<br>*Note: If a $sub\_subcontract$ is being created instead of a $subcontract$, then this step *one* may include a $subordinate\_agent$ deriving a new public sub-key, though a seed value, from its master public key used to sign the parent $subcontract$. All the $master\_agent$, the $master\_contract\_issuer$, the $subordinate\_agent$, and the $subcontract\_issuer$ use the same seed to derive a sub-sub-key or a sub-key, within each one's hierarchy of public keys, from the corresponding parent key. A parent key for the different signatories in this case may be either their master key or their sub-key. |
| two | Depending on the nature of the $subcontract$ being created, the $master\_agent$ either:<br>-uses the location of the $master\_contract$ entry in the repository of contracts; or<br>-embeds a link to the $master\_contract$ entry within the $subcontract$ entry of the repository, and stores the location of the subcontract entry for later use.<br>*Note: the contract repository can be public, private or semi-private, depending on the nature of $contracts$. |
| three | The $master\_agent$ creates a redeem script covering the $subcontract$ being secured, in an $m$-of-$n$ multi-signature structure, where $m$ is the number of compulsory signatures and $n$ further includes the number of non-compulsory signatures and the number of metadata blocks. The number of metadata blocks includes at least the reference to the $master\_contract$ repository entry and the reference to the $subcontract$ entry. Alternatively, this number may include at least a metadata block storing the reference to the $subcontract$ entry where that entry has embedded in it the reference to the $master\_contract$ repository entry, as well as further metadata blocks. |
| four | The $master\_agent$ or the $maste\_contract\_issuer$ pays a nominal amount of crypto-currency to the redeem script calculated in step *three*, through a standard pay-to-script-hash $P2SH$ transaction. |
| five | The $master\_agent$ waits until the $subcontract$ transaction has been published onto the blockchain and extracts its ID. |
| six | *six A*: For a fixed-duration $subcontract$, the $master\_agent$ then creates a new transaction, with a lock-time set to the expiry time of the $subcontract$. This new transaction pays the output from step *five* back to the $master\_agent$ or to the $master\_contract\_issuer$.<br>*six B*: For a $subcontract$ with no fixed duration, the repay script in the new transaction created at step *six* is not time-locked but implementted as an $m$-of-$n$ multi-signature element. This transaction requires a sign-off from a $subordinate\_agent$ monitoring the termination conditions for the $subcontract$, and may be a sign-off from a third party. The multi-signature element may state "subject to sign-off by <x>". The new transaction is then circulated to the required signatories to sign, which include at least <x>. The outputs from such transaction include the fees to <x> and the generated $UTXO$. |

Transactions implementing an example scenario of contract conditionality are shown in Fig. 1. This scenario corresponds to the building contract from Section II.B. The contract includes at least two conditions requiring a planning approval through the issuance of plans certificate and a building-standard approval through the issuance of a final certificate, correspondingly. Building companies often enter in such multiple-counterparty contracts to deliver new buildings. Therefore, it is reasonable to assume that the $contract\ model$ of such contracts already exists, and that there is an entry in the contract repository that can be reused by instantiating it with amended counterparties and parameters. The 'template' contract can be reused simultaneously by several active instantiations, when the building company works in parallel on several projects that target the delivery of different properties. The simultaneous instantiations may also be due to different building companies having an active project each, or having more than one active project each. When a building company reuses the repository entry for the $template\_contract$ for the first time, it creates a new repository entry that acts as the company's own template from then on. That latter template, or rather semi-template, may embed a link to the repository record of the former template. When reusing the semi-template next, the company only appends a line of metadata to the repository entry for that semi-template, and do not create a new repository entry. The appended metadata plays a key role in creating, monitoring and spending crypto-transactions that implement the corresponding instantiation of the contract. The metadata in such transactions include a reference to the company's $semi\_template\_contract$, and a pointer to the line in it containing the specific metadata for this instantiation of the semi-template.

| Representation of the Existence of a New Instance of the Property Building Contract |
|---|
| Transaction-ID: $master\_agent\_T1$ |
| Version number |
| Number of inputs: 1 |
| Previous Transaction Output: <$master\_agent's$ previous unspent BTC output - assume $Y$ Satoshi> |
| Previous Transaction Output Index: IDX-00 |
| Script length |
| ScriptSig: Sig-$master\_agent$ PubK $master\_agent$ |
| Sequence number |
| Number of outputs: 2 |
| First Output value: $Z_1$ <$Z_1$ is less than $Y$> |
| First Output script length |
| First Output script: OP_HASH160 <redeem script hash> OP_EQUAL Redeem Script· requires 2 out of $building\ company$ and $master\_agent$ to conclude: OP_1AssetMetaDataA AssetMetadataB PubK- $master\_agent$ PubK- $building\ company$ OP_4 OP_CHECKMULTISIG**u** |
| Second Output value: $Z_2$ <$Z_1 + Z_2$ is less than $Y$> |
| Second Output length |
| Second Output script: OP_DUP OP_HASH160 <PubK-$master\_agent$ Hash> OP_EQUALVERIFY OP_CHECKSIG |
| LockTime |
| **Creation of a Subcontract by the Master Agent using his first derived key to confirm planning approval** |
| Transaction-ID: $master\_agent\_T2$ |
| Version number |
| Number of inputs: 1 |
| Previous Transaction Output: $master\_agent\_T1$ |
| Previous Transaction Output Index: IDX-01 |
| Script length |
| ScriptSig: Sig-$master\_agent$ PubK $master\_agent$ |
| Sequence number |
| Number of outputs: 2 |
| First Output value: $Z_3$ <$Z_3$ is less than $Z_2$> |
| First Output script length |
| First Output script: OP_HASH160 <redeem script hash> OP_EQUAL Redeem Script requires $building\ control\ department$ to approve and $building\ company$ to approve: OP_2AssetMetaDataA AssetMetadataB PubK-$master\_agent$ SK1 PubK-$building\ control\ department$ PubK-$building\ company$ OP_5 |
| Second Output value: $Z_4$ <$Z_3 + Z_4$ is less than $Z_2$> |
| Second Output length |
| Second Output script: OP_DUP OP_HASH160 <PubK-$master\_agent$ Hash> OP_EQUALVERIFY OP_CHECKSIG |
| LockTime |
| **Creation of a Subcontract by the Master Agent using his second derived key to confirm building-standard approval** |
| Transaction-ID: $master\_agent\_T3$ |
| Version number |
| Number of inputs: 1 |
| Previous Transaction Output: $master\_agent\_T2$ |
| Previous Transaction Output Index: IDX-01 |
| Script length |
| ScriptSig: Sig-$master\ agent$ PubK $master\_agent$ |
| Sequence number |
| Number of outputs: 2 |
| First Output value: $Z_5$ <$Z_5$ is less than $Z_4$> |
| First Output script length |
| First Output script: OP_HASH160 <redeem script hash> OP_EQUAL Redeem Script requires $approved\ inspector$ to approve and $building\ company$ to approve: OP_2AssetMetaDataA AssetMetadataB PubK-$master\_agent$SK2 PubK-$approved\ inspector$ PubK-$building\ company$ OP_5 |
| Second Output value: $Z_6$ <$Z_5 + Z_6$ is less than $Z_4$> |
| Second Output length |
| Second Output script: OP_DUP OP_HASH160 <PubK-$master\_agent$ Hash> OP_EQUALVERIFY OP_CHECKSIG |
| LockTime |
| **Planning Authority Sign-off** |
| Transaction-ID: $master\_agent\_T4$ |
| Version number |
| Number of inputs: 1 |
| Previous Transaction Output: $master\_agent\_T2$ |
| Previous Transaction Output Index: IDX-00 |
| Script length |
| ScriptSig: Sig-$building\ control\ department$ Sig-$building\ company$ OP_2AssetMetaDataA AssetMetadataB PubK- $master\_agent$SK1 PubK- $building\ control\ department$ PubK-$building\ company$ OP_5 OP_CHECKMULTISIG |
| Sequence number |
| Number of outputs: 1 |
| Output value: $Z_7$ <$Z_7$ is less than $Z_3$> |
| Output script length |
| Output script: OP_DUP OP_HASH160<$building\ control\ department$ Hash>OP_EQUALVERIFY OP_CHECKSIG <The $building\ control\ department$ is paid a fee in Satoshi> |
| LockTime |

Figure 1. Creating crypto-transactions corresponding to contract and subcontract start, execution and completion.

Within the automated management of a building company's $semi\_template\_contract$, a $master\_agent$ associated with the semi-template monitors for a new line of parameters being appended. New lines are appended by the building company. The mechanism involves the company routinely allocating some amount of crypto-currency to the $master\_agent$, so that the agent can activate at any time the first steps in its algorithm on issuing a new instance of the contract. The first step in that algorithm is the creation and broadcast of the transaction shown in dark shade in Fig. 1, and extracting its ID $T_1$ after the transaction is recorded on the blockchain. Thus, $T_1$ becomes a secure, immutable, and publically available electronic record of the existence of the new contract in the physical world. The amount of crypto-currency $Y$ accessed for this step by the $master\_agent$ can be small. The amount allocated by the building company for access by the master agent is reviewed at routine intervals. At the end of an interval, the balance (excluding a set minimum) is automatically returned to the building company, and it is assessed if and what amount to make available to the agent in the next period. When the activity of a company is more versatile and it is captured through several semi-templates of different type, then the allocation and reallocation of crypto-currency amounts to the $master\_agents$ of the templates is managed and optimized by a higher-hierarchy agent called $templates\_manager$. The reallocation is based on the evaluation of the prevailing performance and usual needs of the $master\_agents$. That performance and needs are linked to the type of business line a semi-template is supporting within the company's business portfolio, and the performance of the company along the different business lines.

Following $T_1$, Fig. 1 presents next (in light grey) transactions related to two of the subcontracts that the $master\_agent$ of this template manages within an instantiation of the $template\_contract$. First, $T_2$ indicates that a $subcontract_1$ for getting a planning approval exists, and next, $T_3$ indicates that a $subcontract_2$ exists for getting a building-standard approval. On the other hand, transaction $T_4$ confirms that a planning approval is received, and pays the fees to the local authority's building control department. Therefore, the first $subcontract_1$ is now closed. The complete conditionality structure of a building contract is more complex, involves a larger number of $subcontracts$, and in some cases also involves $sub\_subcontracts$ and $subordinate\_agents$. The hierarchical structure emerges in Fig. 1, and shows that the $master\_agent$ derives a secret (private) sub-key $SK_1$ for managing $subcontract_1$ and a secret sub-key $SK_2$ for managing $subcontract_2$. Section III next discusses hierarchical structures in more detail.

## III. SMART CONTRACTS' EFFICIENCY AND SECURITY

### A. Hierarchical Structures of Contracts, Crypto-keys, and Common Secrets abeted

The automated management of smart contracts [4] introduced in Sections II contributes to scaling blockchain functionality. Section II.C indicates that managing contract conditionality is assisted by a hierarchical structure of public/private key-pairs. The derivation of crypto-key hierarchies [5][11] supports the execution and management of smart contract. Let us consider a complex contract's conditionality implemented through a hierarchy of $subcontracts$ and $sub\_subcontracts$, and assisted through an hierarchy of sub-keys and sub-sub-keys. Fig. 2 shows a tree structure in blue representing the hierarchical contract conditionality, and a corresponding three structure in red representing hierarchical keys needed to assist the implementation of the complex contract. Each element in the red tree corresponds to a public/private key-pair, created by adding multi-rehashed relevant information. That information may include IDs of existing transactions or metadata from existing entries in the contract repository. Though only the secret (private) keys $SK$ are indicated in the red tree, a corresponding public key $PK$ is also derived for every derived secret key. Therefore, the tree corresponds to a hierarchy of asymmetric public/private key-pairs $PK/SK$. For clarity of introducing the mechanism, it is assumed that the $master\_agent$ $MA$ manages all $subcontracts$ and $sub\_subcontracts$. In practice, some of these elements of the blue structure can be managed by $subordinate\_agents$.

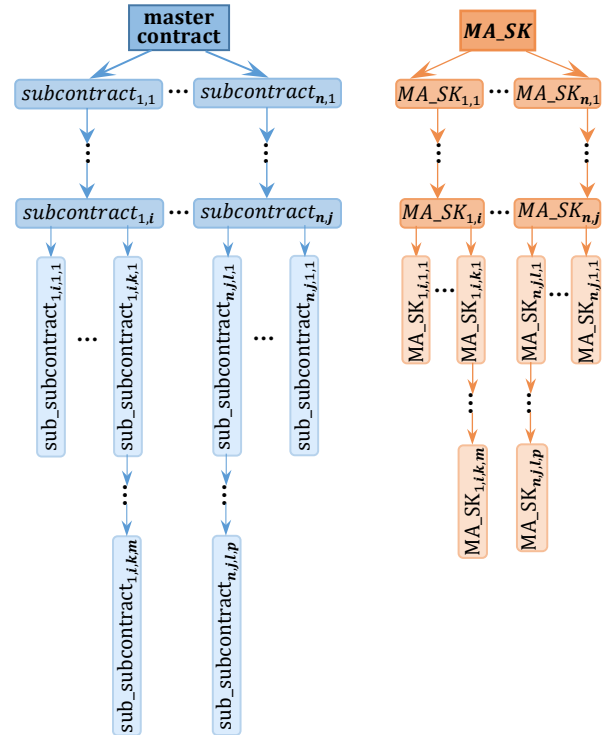

Figure 2. Hierarchical contract conditionality and corresponding derived tree-structure of keys.

Notice that each element of the blue structure is implemented by at least two transactions, i.e. indicating that a new (sub-sub-)contract exists and then terminating it. This is the case with transactions $T_2$ and $T_4$ in Fig. 1, for example. Therefore, the pair in each element of the red structure is used to sign and redeem scripts in at least two transactions. Next, the notation $MA\_SK$ in Fig. 2 refers to the master private key of the $master\_agent$ ($MA$), and $MA\_SK_{1,1}$ to $MA\_SK_{n,1}$ are private sub-keys of the same agent, where $subcontract_{1,1}$ to $subcontract_{n,1}$ can be executed in parallel. On the other hand, $MA\_SK_{1,1}$ and $MA\_SK_{1,i}$ are the private sub-keys of this agent, where $subcontract_{1,1}$ to $subcontract_{1,i}$ can only be executed in sequence. Further, $MA\_SK_{1,i,1,1}$ to $MA\_SK_{1,i,k,1}$ are $MA$'s private sub-sub-keys, where $sub\_subcontract_{1,i,1,1}$ to $sub\_subcontract_{1,i,k,1}$ can be executed in parallel only after $subcontract_{1,1}$ to $subcontract_{1,i}$ are executed in sequence. Finally, $MA\_SK_{n,j,1,1}$ to $MA\_SK_{n,j,l,p}$ are $MA$'s private sub-sub-keys, where $sub\_subcontract_{n,j,1,1}$ to $sub\_subcontract_{n,j,l,p}$ can only be executed in sequence and only after $subcontract_{n,1}$ to $subcontract_{n,j}$ are executed in sequence. Notice that some $subcontracts$ can serve as $master\_contracts$ for the $sub\_subcontracts$ that follow below them in the hierarchical structure. Thus, $subcontract_{1,i}$ can act as a $master\_contract$ or rather as a $submaster\_contract$ for $sub\_subcontract_{1,i,1,1}$ to $sub\_subcontract_{1,i,k,1}$ as well as for $sub\_subcontract_{1,i,k,1}$ to $ub\_subcontract_{1,i,k,m}$.

In the general case, any information can serve the role of a seed. However, the information may also be meaningful in the contexts that the hierarchy of keys is used. We choose here as $aster\_seed$ $M$, the redeem script hash securing the $master\_contract$ and created in an $m$-of-$n$ multi-signature structure. Further, a $sub\_master$ seed $SM$ is chosen as the redeem script hash securing a $sub\_master\_contract$ and created in an $m$-of-$n$ multi-signature structure. From Fig. 2, it can be deducted that at least sub-master seeds $SM_{1,i}$ and $SM_{n,j}$ must be chosen, corresponding to $submaster\_contract_{1,i}$ and $sub\_master\_contract_{n,j}$. The seeds $M$, $SM_{1,i}$, $SM_{n,j}$ are involved in producing the generator values from $GV_{1,1}$ to $GV_{n,j,l,p}$, when deriving the tree of asymmetric cryptographic key-pairs. Key derivation starts with the $master\_agent$ selecting a random value for the base point $B$, and communicating it to the $contract\_issuer$. The base point is applied, as described below, to derive a public key from a corresponding private key, in order to complete an asymmetric cryptographic key-pair using Elliptic Curve Cryptography (ECC). The base point can also be communicated to any other signees on transactions created in implementing the hierarchical contract conditionality, particularly if they have a significant role in the structure and that role involves communications/negotiations in relation to a number of elements in the structure. We will introduce the mechanism first focusing on the $master\_agent$ and the $contruct\_issuer$, and their meaningful hierarchies of cryptographic key-pairs and $common\_secrets$. However, this mechanism can be applied accordingly when other signees also derive their hierarchies of key-pairs and $common\_secrets$. The mechanism can further be adapted to the case when some branches of the conditionality structure are managed by $submaster\_agents$.

Considering Fig. 2, the $master\_agent$ $MA$ starts with its ECDSA-valid [12] secret key $MA\_SK$, where ECDSA abbreviates the Elliptic Curve Digital Signature Algorithm. Then, $MA$ derives its hierarchy of private keys as follows:

$$\left. \begin{aligned} MA\_SK_{r,1} &= MA\_SK + GV_{r,1} \\ GV_{r,1} &= SHA\_256(M, L_{r,1}) \end{aligned} \right\} \text{ for } 1 \leq r \leq n \quad (1)$$

$$MA\_SK_{1,r} = MA\_SK_{1,r-1} + SHA\_256^{r-1}(M) \quad (2)$$
$$\text{for } 2 \leq r \leq i$$

$$\left. \begin{aligned} MA\_SK_{1,i,r,1} &= MA\_SK_{1,i} + GV_{1,i,r,1} \\ GV_{1,i,r,1} &= SHA\_256(SM_{1,i}, L_{1,i,1,1}) \end{aligned} \right\} \text{ for } 1 \leq r \leq k \quad (3)$$

$$MA\_SK_{n,j,l,r} = MA\_SK_{n,j,l,r-1} + SHA\_256^{r-1}(SM_{n,j}) \quad (4)$$
$$\text{for } 2 \leq r \leq p$$

where

$$SHA\_256^r(M) = SHA\_256(SHA\_256^{r-1}(M)) \quad (5)$$

For $1 \leq r \leq n$, generator value $GV_{r,1}$ is produced using the concatenation $(M, L_{r,1})$ of the redeem script hash M of the $master\_contract$ and the $template\_contract$ hash $L_{r,1}$. Here, $subcontract_{r,1}$ is an instantiation of the $semi\_template\_contract$ which is being amended from the $template\_contract_{r,1}$. Further, for $1 \leq r \leq k$, generator values $GV_{1,i,r,1}$ use the $submaster\_seed$ $SM_{1,i}$ instead of the $master\_seed$ $M$. Also, $GV_{1,i,r,1}$ are produced by analogy to $GV_{r,1}$, as $sub\_subcontracts_{1,i,r,1}$ for $1 \leq r \leq n$ are executed in parallel, similarly to the way $sub\_subcontracts_{r,1}$ for $1 \leq r \leq k$ are executed in parallel. On the other hand, generator values $GV_{1,r}$ are produced by rehashing the $master\_seed$ $M$ for $2 \leq r \leq i$, and values $GV_{n,j,l,r}$ are produced by rehashing the $submaster\_seed$ $SM_{n,j}$ for $2 \leq r \leq p$.

Next, Elliptic Curve Cryptography (ECC), properties of elliptic curve operations, and the base point $B$ are used to complete the asymmetric cryptographic key-pairs and derive the public keys of the $master\_agent$. The operator + in (6-10) stands for scalar addition and the operator × refers to elliptic curve point multiplication. Having that elliptic curve cryptography algebra is distributive, the hierarchy of public keys of the $master\_agent$ is produced as follows:

$$MA\_PK = MA\_SK \times B \quad (6)$$

$$MA\_PK_{r,1} = MA\_PK + GV_{r,1} \times B, \text{ for } 1 \leq r \leq n \quad (7)$$

$$MA\_PK_{1,r} = MA\_PK_{1,r-1} + SHA\_256^{r-1}(M) \times B \quad (8)$$
$$\text{for } 2 \leq r \leq i$$

$$MA\_PK_{1,i,r,1} = MA\_PK_{1,i} + GV_{1,i,r,1} \times B \text{ for } 1 \leq r \leq k \quad (9)$$

$$MA\_PK_{n,j,l,r} = MA\_PK_{n,j,l,r-1} + SHA\_256^{r-1}(SM_{n,j}) \quad (10)$$
$$\text{for } 2 \leq r \leq p$$

where $GV_{r,1}$ are the generator values used in (1) and $GV_{1,i,r,1}$ are the generator values used in (3). By analogy with the $master\_agent's$ hierarchy of key-pairs, the public/private key pairs of the $contract\_issuer$ CI are derived as follows:

$$CI\_SK_{r,1} = CI\_SK + GV_{r,1} \text{ for } 1 \leq r \leq n \quad (11)$$

$$CI\_SK_{1,r} = CI\_SK_{1,r-1} + SHA\_256^{r-1}(M) \quad (12)$$
$$\text{for } 2 \leq r \leq i$$

$$CI\_SK_{1,i,r,1} = CI\_SK_{1,i} + GV_{1,i,r,1} \text{ , for } 1 \leq r \leq k \quad (13)$$

$$CI\_SK_{n,j,1,r} = CI\_SK_{n,j,1,r-1} + SHA\_256^{r-1}(SM_{n,j}) \quad (14)$$
$$\text{for } 2 \leq r \leq p$$

$$CI\_PK = CI\_SK \times B \quad (15)$$

$$CI\_PK_{r,1} = CI\_PK + GV_{r,1} \times B \text{ , for } 1 \leq r \leq n \quad (16)$$

$$CI\_PK_{1,r} = CI\_PK_{1,r-1} + SHA\_256^{r-1}(M) \times B \quad (17)$$
$$\text{for } 2 \leq r \leq i$$

$$CI\_PK_{1,i,r,1} = CI\_PK_{1,i} + GV_{1,i,r,1} \times B \text{ for } 1 \leq r \leq k \quad (18)$$

$$CI\_PK_{n,j,l,r} = CI\_PK_{n,j,l,r-1} + SHA\_256^{r-1}(SM_{n,j}) \quad (19)$$
$$\text{for } 2 \leq r \leq p$$

In scripts and transactions related to different $subcontracts$ or $sub\_subcontracts$, the $master\_agent$ and the $contract\_issuer$ use different corresponding keys within each one's hierarchy of keys. This increases security, as even if a transaction related to a $(sub\_)subcontract$ is compromised, the integrity of the rest of the smart contract structure is preserved. Furthermore, an element in one's hierarchy of public keys can be produced in advance of the execution of the corresponding $(sub\_)subcontract$, as the relevant generator value is available and known to the $master\_agent$ and the $contract\_issuer$ before that execution. Notice that the generator values can be produced in a different way, and (1-19) present just one alternative. However, any version should allow the evaluation of the current generator value before the current element of the conditionality structure. Thus, the $master\_agent$ evaluates each of the public keys $CI\_PK_{r,1}$ to $CI\_PK_{n,j,1,r}$ of the $contract\_issuer$ at the same time at which the $contract\_issuer$ evaluates them. The vice-verse is also true, and the $contract\_issuer$ evaluates each of the public keys $MA\_PK_{r,1}$ to $MA\_PK_{n,j,1,r}$ of the $master\_agent$ at the same time at which the $master\_agent$ evaluates them. The pairing private keys are produced by their owner also at that same time, i.e. the earliest step he can produce the corresponding public keys.

The produced hierarchies of cryptographic keys can also be used to produce an hierarchy of encryption keys. Notice that the $contract\_issuer$ and the $master\_agent$ produce independently the same hierarchy of $common\_sercrets$ CS, as presented in Fig. 3. The $contract\_issuer$ produces the $common\_secrets$ as:

$$CS_{r,1} = CI\_SK_{r,1} \times MA\_PK_{r,1} \text{ , for } 1 \leq r \leq n$$
$$\vdots \quad (20)$$
$$CS_{n,j,l,r} = CI\_SK_{n,j,l,r} \times MA\_PK_{n,j,l,r} \text{ , for } 2 \leq r \leq p$$

and the $master\_agent$ produces the same $common\_secrets$ as:

$$CS_{r,1} = MA\_SK_{r,1} \times CI\_PK_{r,1} \text{ , for } 1 \leq r \leq n$$
$$\vdots \quad (21)$$
$$CS_{n,j,l,r} = MA\_SK_{n,j,l,r} \times CI\_PK_{n,l,1,r} \text{ , for } 2 \leq r \leq p$$

Now, each $common\_secret$ $CS_{r,1}$ to $CS_{n,j,l,r}$ serves as a basis for a symmetric encryption key securing a channel for communication between $CI$ and $MA$ regarding a corresponding $subcontract_{r,1}$ to $sub\_subcontract_{n,j,l,r}$. For example, the communication may confirm parameters or prioritize preferences.

### B. Efficient and Secure Transfer of Smart Contracts and Underlying Entities

Smart contract functionality can be extended further by: (i) introducing mechanisms for efficient and secure transfer of entities on a blockchain [6][7], and then (ii) using these mechanisms in cases where the entities being transferred are underlying smart contracts. Notice that a smart contract can also be an underlying of another smart contract. For example, the ownership of a tokenized financial instrument is transferred through a smart contract, and the structure of the financial instrument is implemented through another smart contract. The underlyings can include physical assets and IoT devices manipulated through the contracts, or virtual assets – such as rights on physical assets or rights on particular services – that are controlled through the contract. The increased smart-contract functionality, in turn, helps sustain blockchain-enabled services of varying complexity.

The transfer of entities underlying smart contracts is facilitated trough tokenization techniques. Enhanced optimization of memory usage in the electronic transfers, and improved security and data integrity are achieved through hashing techniques. Steps in the transfer mechanism involve:
- Generating a script $S_k$ that comprises:
  - A set of metadata $D_k$ associated with an invitation for the exchange of an entity $E_k$, where $E_k$ is one of the underlyings of a smart $master\_contract$ or

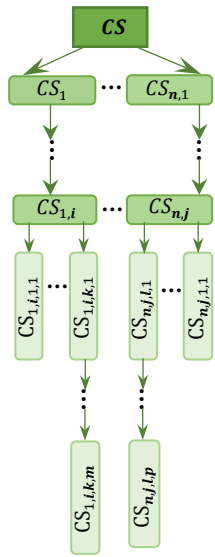

Figure 3. A hierarchical of common secrets.

*subcontract*. The meta-data includes a pointer or other reference to the location of that contract.
- The derived public key $CI_A\_PK_k$ associated with the $contract\_issuer_A$ and used in scripts, transactions and communication in relation to the exchange of entity $E_k$ owned by $A$.
- The derived public key $MA_A\_PK_k$ of the $master\_agent_A$ managing the contract issued by $A$, where $MA_A\_PK_k$ is used only in relation to entity $E_k$. (depending on the overall contract's conditionality structure, the agent here can be a $subordinate\_agent_A$, as well)

- Hashing $S_k$ and publishing $S_k$ and its hash on a distributed hash table (DHT), which is distributed across a (worldwide) network and the script hash serves as a DHT look-up key.
  - This DHT resource differs from the DHT repository of contracts discussed in Section II.
- Generating an invitation transaction $TI_k$ for inclusion on the blockchain, where the transaction comprises the hash of $S_k$ and an indication of an entity $E'_k$ to be transferred in exchange for $E_k$.
- Scanning through the plurality of DHT entries, where each entry comprises:
  - an invitation to perform an exchange of an entity $E_n$ underlying a smart contract; and
  - a link to an invitation transaction $TI_n$ on the blockchain.
- (Partial) matching of the set of metadata $D_k$ from the initial invitation-entry in the DHT repository of invitations, to a set of metadata $D_m$ in another invitation-entry. Each set $D_k$ and $D_m$ comprises:
  - an indication of entities to be exchanged, $E_k$ for $E'_k$ and $E_m$ for $E'_m$, correspondingly, where $E'_k \approx E_m$ and $E'_m \approx E_k$, and
  - conditions for the exchange that also (partially) match.
- Generating, broadcasting, and recording on the blockchain of an exchange transaction $TE_{k,m}$ that includes:
  - The script $S_k$, signed with the derived private key $MA_A\_SK_k$ of the $master\_agent_A$, where $MA_A\_SK_k$ corresponds as cryptographic pairing to the public key $MA_A\_PK_k$. The script $S_k$ may also be signed by the $contract\_issuer_A$ using the private key $CI_A\_PK_k$. (the agent here can be a $subordinate\_agent_A$, as well)
  - The script $S_m$ of the (partially) matching DHT invitation-entry, signed with the private key $MA_B\_SK_m$ corresponding to public key $MA_B\_PK_m$, and signed with the private key $CI_B\_SK_m$ corresponding to $CI_B\_PK_m$. These keys are associated, respectively, with the $master\_agent_B$ and $contract\_issuer_B$. (the agent here can be a $subordinate\_agent_B$, as well)
  - A first input provided from an output of invitation transaction $TI_k$.
  - A second input provided from an output of invitation transaction $TI_m$.
  - A first output indicating a quantity of the (tokenized) entity $E_k$ to be transferred to the control of $master\_contract_B$, and to the ownership of $contract\_issuer_B$. (the smart contract here can be a $subcontract_B$, as well)
  - A second output indicating a quantity of the (tokenized) entity $E_m$ to be transferred to the control of $smart\_contract_A$ and to the ownership of $onntract\_issuer_A$. (the smart contract here can be a $subcontract_A$, as well)

In a hypothetical example, pension funds offer a variety of structured pension products and clients can hold a portfolio of different structured products from different funds. For each of the structured pension products a client holds, he may also select the proportion of the elements within the product. Clients of different funds are allowed to exchange (parts of) their holdings under certain conditions. The conditions differ among funds in level of detail and restrictive constraints, and so the exchange is not standardized. In the context of pensions, the actions are of relatively low frequency and based on long term perspective. When a client $C$ would like to exchange parts of his holdings, he acts as a $contract\_issuer_C$. The smart contract managing the exchange of all structured pension products ($P$) he holds is a $subcontract_C^P$ within a $master\_contract_C$ managing his financial assets. As a $contract\_issuer_C$ ($CI_C$), the client $C$ has different public/private pairs of cryptographic sub-keys derived from his master pair of keys $CI_C\_PK/CI_C\_SK$. One of these pairs of sub-keys, $CI_C\_PK^P/CI_C\_SK^P$, is associated with the $subcontract_C^P$. The $master\_contract_C$ and the $subcontract_C^P$ are managed, correspondingly, by a $master\_agent_C$ and a $subordinate\_agent_C^P$. The $master\_agent_C$ has its own pair of master keys $MA_C\_PK/MA_C\_SK$ and the derived from them pairs of sub-keys. One of these sub-key pairs, $MA_C\_PK^P/MA_C\_SK^P$, is associated with the client's pension holdings. The $subordinate\_agent_C^P$ ($SA_C^P$) has a pair of cryptographic keys $SA_C^P\_PK/SA_C^P\_SK$, and uses them for the transactions associated with the start and closure of the $subcontract_C^P$. Following the transfer mechanism described in this Section III.B, each one of the pension products held by the client $C$ is an entity $E_k$ underlying the $subcontract_C^P$ and this subcontract has $r$ underlying entities, $1 \leq k \leq r$. The exchange of each entity $E_k$ is further associated with a separate pair of cryptographic keys $SA_C^P\_PK_k/SA_C^P\_SK_k$ used by $SA_C^P$ within the same $subcontract_C^P$. These keys are derived from $SA_C^P\_PK/SA_C^P\_SK$ using the algorithm from Eqs. (1-10). The generator values in this algorithm are now based on information about entries in the (worldwide) DHT exchange repository of entities underplaying smart contracts. Some of the entities $E_k$, $1 \leq k \leq r$, underlying the same $subcontract_C^P$, can be exchanged in parallel and others in sequence, depending on restrictions by different pension funds. Each pair $SA_C^P\_PK_k/SA_C^P\_SK_k$ is used with script $S_k^P$, invitation transaction $TI_k^P$, and exchange transaction $TE_{k,m}^P$, for $1 \leq k \leq r$. For each $k$, a common secret $CS_k^P$ is produced from the cryptographic key-pairs $SA_C^P\_PK_k / SA_C^P\_SK_k$ and $CI_C\_PK_k^P / CI_C\_SK_k^P$, using the algorithm from Eqs. (11-21). The encryption key based on the common secret $CS_k^P$ is used to provide a secure channel of communication about the corresponding structured pension product $E_k$. Notice, as well, that each $E_k$ a client holds is itself a smart contract managing the constitution of a structured financial product, and so the entities underlying $subcontract_C^P$ are smart contracts. These contracts $E_k$ are not sub-subcontracts of $subcontract_C^P$ and are not in its conditionality structure. Each contract $E_k$ implements the logic of the corresponding financial instruments. $E_k$ is not necessarily a tokenization of an existing off-chain instrument but also a creation of a new financial instrument that does not exist off-chain.

The method proposed in this Section III.B provides for data integrity and optimization of memory. The DHT invitation-entries, initiated at different stages of the conditionality structures of a variety of smart contracts, may be matched worldwide or within a scope indicated in / required by the smart contracts. The method enables disparate smart contracts and subcontracts to identify and match each other in terms of underlying entities, and to securely exchange these underlyings. The method does not require alteration of existing blockchain protocols, while embedding metadata in scripts associated with blockchain transactions.

## IV. CONCLUSION

This paper introduces methods for extending the functionality of blockchain-enforced smart contracts. We propose a mechanism for automated management of smart contracts with hierarchical conditionality structures, and a mechanism for efficient and secure matching and transfer of contract underlyings among diverse smart contracts and subcontracts. Services, enabled by implementing blockchain-enforced smart contracts with extended functionality, are secure, efficiently automated, and allow (worldwide) resource distribution. They present sustainable alternatives and complements to some of the current service infrastructures within a range of domains., particularly when some of them are underperforming or with unreliable security.

Our research focus is next on big-data analysis of the potential effects on the performance of blockchain networks and services due to: (i) adopting different innovative methods for extending blockchain functionality, (ii) the rate of adoption, and (iii) the sequence in which they are adopted.

ACKNOWLEDGMENT

The authors are grateful for the creative and supportive environment at nChain Limited, within its scientific and applied research teams.